\documentclass[letterpaper,twoside]{jpconf}
\usepackage{graphicx}
\usepackage{latexsym}
\usepackage{hyperref}
\newcommand{\slj}[3]{\mbox{$^{#1}${\ifcase#2\or S\or 
	 P\or D\or F\or G\fi}$_{#3}$}}
\newcommand{\sLj}[3]{{}^{#1}\!#2_{#3}}
\newcommand{\jpsi}{\ensuremath{J\!/\!\psi}}

\newcommand{\gev}{\hbox{ GeV}}
\newcommand{\mev}{\hbox{ MeV}}

\newcommand{\cfrac}[2]{\textstyle \frac{#1}{#2}}
\catcode`\@=11
\def\ps@fnal{\def\@oddhead{\textsf{FERMILAB--Conf--04/317--T \hfil \thepage}}
\def\@evenhead{\thepage \hfil \textsf{FERMILAB--Conf--04/317--T}}}
\relax
\begin{document}
\pagestyle{fnal} 
\phantom{M} \hfill \textsf{FERMILAB--Conf--04/317--T}\\[-24pt]

\title{Theoretical Overview: The New Mesons}

\author{Chris Quigg}

\address{Theoretical Physics Department, Fermi National Accelerator 
Laboratory,\\ P.O.~Box 500, Batavia, Illinois 60510 USA}

\ead{quigg@fnal.gov}

\begin{abstract}
    After commenting on the state of contemporary hadronic physics and 
    spectroscopy, I highlight four areas where the action is: 
    searching for the relevant degrees of freedom, mesons with beauty 
    and charm, chiral symmetry and the $D_{sJ}$ levels, and $X(3872)$ 
    and the lost tribes of charmonium. 
\end{abstract}
\section{Introduction \label{sec:intro}}
\subsection{Three Cheers for QCD! \label{subsec:threec}}
It has been a good season for Quantum Chromodynamics. Just a few days 
ago, the Royal Swedish Academy of Sciences awarded the 2004 Nobel 
Prize in Physics to  David Gross, David Politzer, and 
Frank Wilczek, ``for the discovery of asymptotic freedom in the theory 
of the strong interaction.'' I hope that all of you have not only 
taken pleasure in the recognition accorded to our friends and their 
work~\cite{Gross:1973id,Politzer:1973fx}, but also have taken some time to reflect on the developments 
that preceded and followed from the discovery of asymptotic 
freedom,\footnote{A good starting place is the informal history to be 
found at \url{http://nobelprize.org/physics/laureates/2004/phyadv04.pdf}.} 
and to share your understanding of the remarkable edifice that is QCD with 
your colleagues at home. 

The discovery of asymptotic freedom and the understanding of what it 
could mean marked a tipping point in the long struggle to make sense 
of the strong interactions, but it could only have appeared in a rich 
context. If by chance you have never read the seminal papers, or if it
has been a long time since you enjoyed them, let me point you to a 
short selection of the classics. The idea of a vector gluon theory 
may be found in a prescient paper by Nambu~\cite{Nambu}, and the case 
for a color gauge theory is made by 
Gell-Mann and collaborators~\cite{Murray,Bardeen:1972xk,Fritzsch:1973pi}. Perceiving the broad reach of the 
antiscreening property of non-Abelian gauge theories required an 
understanding of the 
Callan~\cite{Callan:1970yg} -- Symanzik~\cite{Symanzik:1970rt} 
renormalization group equations, which many physicists of a certain 
age acquired from Sidney Coleman's Erice 
lecture notes~\cite{Sidney}. And, of course, the tension that 
asymptotic freedom relieved was created by the successes of the quark 
model, the unavailing searches for free quarks, the experimental 
observation~\cite{Friedman:1972sy} of Bjorken scaling~\cite{Bjorken:1968dy}, and 
the intuitive appeal of the parton model~\cite{Bjorken:1969ja,RPF}. 
Now is the perfect time to refresh your appreciation of this wonderful 
saga, and to look ahead to what else we might make.

\subsection{Spectroscopy Is Alive and Well \label{subsec:spectroren}}
The occasion to celebrate past glories isn't the only reason to see 
this as a good season for QCD. The talks we have just heard by Gabriella 
Sciolla~\cite{gab}  and Steve 
Olsen~\cite{steve}  are emblematic of 
a most remarkable renaissance in hadron spectroscopy now in full bloom.
Over the past few
years, experiments have uncovered a plethora of new narrow states that
extend our knowledge of hadrons and challenge our understanding of the
strong interaction.  

In the heavy-flavor arena that I shall emphasize, 
first came the discovery in the Belle experiment
of $\eta_{c}^{\prime}$ in exclusive $B \to K K_{S} K^{-}\pi^{+}$
decays~\cite{Choi:2002na}.  CLEO~\cite{Asner:2003wv},
BaBar~\cite{Wagner:2003qb}, and Belle~\cite{Abe:2003ja} have confirmed
and refined the discovery of $\eta_{c}^{\prime}$ in $\gamma\gamma$
collisions, fixing its mass and width as $M(\eta_{c}^{\prime}) = 3637.7
\pm 4.4 \mev$ and $\Gamma(\eta_{c}^{\prime}) = 19 \pm
10\mev$~\cite{Skwarnicki:2003wn}.  The unexpectedly narrow $D_{sJ}$
states discovered by Babar~\cite{Aubert:2003fg},
CLEO~\cite{Besson:2003cp}, and Belle~\cite{Abe:2003jk} provided the next
surprise.  Then followed the discovery by Belle~\cite{Choi:2003ue} of
$X(3872) \to \pi^+\pi^-\jpsi$, rapidly confirmed by
CDF~\cite{Acosta:2003zx}, D\O~\cite{Abazov:2004kp}, and 
BaBar~\cite{Aubert:2004ns}.   This year, it appears that we may 
finally have solid evidence for $h_{c}$, the 1\slj{1}{2}{1} level of 
charmonium, from E835~\cite{Patrignani} and CLEO~\cite{kam,tomaradze}. 
We have just heard~\cite{steve} that Belle has sighted one or more 
interesting peaks in the neighborhood of $3940\mev$. On the baryon 
side, SELEX has reported signals for a family of doubly charmed 
baryons, beginning with $\Xi_{cc}^{+}(3519)$~\cite{Mattson:2002vu}. 
And the broad $j_{q} = \cfrac{1}{2}$ charm 
levels are coming under increasing experimental 
scrutiny~\cite{Anderson:1999wn,Link:2003bd,Abe:2003zm,Kutschke:2004wq}.

In the light-quark sector, evidence has been accumulating for exotic 
$J^{PC} = 1^{-+}$ mesons~\cite{Adams:1998ff,adams} that do not 
correspond to $q\bar{q}$ states.
Conflicting evidence on pentaquarks,  baryon states with 
quantum numbers that do not occur in the simple $qqq$ quark-model
description of baryons, continues to provoke lively 
discussion~\cite{hicks,dzierba,litvintsev,qiang}.    

Each of the
new states raises questions of interpretation, and offers
opportunities to challenge---and deepen---our understanding. The 
conversation between theory and experiment, and among different 
experiments, has been lively and stimulating.


\subsection{The Value of Hadronic Physics \label{subsec:helen}}
It is good to engage in lively conversations; it is even better to 
reach significant conclusions about the way Nature works. And there 
the exploratory---and, not infrequently, wooly---character of hadronic 
physics means that the path to fundamental lessons may be indistinct. 
The hadronic physics community needs to think about how it spends its 
time---and why.

In her opening lecture at \textit{Heavy Quarks and Leptons 2004}, Helen
Quinn~\cite{Quinn:2004qb} made the following observations to explain
her emphasis on the weak interactions: 
\begin{quote}
    Hadron phenomenology and spectroscopy does not test the standard
    model.  We have a qualitative understanding of QCD phenomenology,
    but many aspects are not calculable from first principles.  We make
    models for new states: approximations such as potential models, or
    intuitive pictures of substructure.  The competing pictures are not
    mutually exclusive; quantum superpositions are possible.  We'll
    never throw out QCD if these pictures turn out not to work for the
    next state we find.  We do learn how to refine our approximations
    to QCD.
\end{quote}
These are fair and thoughtful observations. How do you respond?  

I would note that there is value to both fundamental and applied
science, and that the apparently less glamourous work of applied
science may be just what we need to get at the fundamental lessons. 
Moreover, exploration---the task of discovering what phenomena exist 
and of developing systematics---helps us to understand what the 
fundamental questions are, and how we might best address them.

At the same time, a little self-awareness (even self-criticism) can 
make for better science. We need to distinguish between exploratory 
models and controlled approximations, to pay attention to whether we 
are testing theories or theorists (both activities have some social 
value), and to ask whether we are probing close to the Lagrangian of 
Quantum Chromodynamics or comparing random numbers with experiment.

Physics doesn't advance by perturbation theory alone, and it is worth
recalling that one of QCD's signal achievements is explaining what sets
the mass of the proton---or, if you like, what accounts for nearly all
the visible mass of the Universe.  The insight that the mass of the 
proton arises from the energy stored up in confining three quarks in a 
small volume, not from the masses of the constituents themselves, is a 
landmark in our understanding of Nature~\cite{Wilczek:1999be}. The 
value of that insight isn't diminished because it is a little bit 
qualitative, or because a quantitative execution of the idea requires 
the heavy machinery of lattice field theory.

More generally, there is great value in a convincing physical picture
that can show us the way to an answer (whether or not precise and
controlled), or show that some tempting simplifying assumptions are
unwarranted.  The chiral quark model~\cite{Manohar:1983md}, which
identifies the significant degrees of freedom on the 1-GeV scale as
constituent quarks and Goldstone bosons, offers a nice example.  It
points to the $u$-$d$ asymmetry in the light-quark sea of the
proton~\cite{Eichten:1991mt}, and predicts a negative polarization of
the strange (but not antistrange) sea. A lifetime of staring at 
$\mathcal{L}_{\mathrm{QCD}}$ wouldn't lead to these expectations.

We can value \textit{anschaulich} explanations as sources of intuition
and instruments of exploration, while keeping clearly in mind their
limitations, as we try to address many open-ended questions, including:
What is a hadron?  What are the apt degrees of freedom?  What
symmetries are fruitful?  What are the implications of QCD under
extreme conditions?

\subsection{The Theory of Everything \label{subsec:emerge}}
There is another reason to contemplate the style of analysis that
characterizes hadronic physics and its relationship to fundamental 
insights.  In a provocative article, Laughlin \&
Pines~\cite{Laugh} trumpet the end of reductionism (``the science of
the past,'' which they identify with particle physics) and the triumph
of emergent behavior, the study of complex adaptive systems (``the
physics of the next century''). By \textit{emergent,} we understand 
phenomena that are not simply derived from the underlying 
microphysics---a Lagrangian, say---but are governed by ``higher
organizing principles'' (perhaps universal), relatively independent of
the fundamental theory. If these clever fellows were better listeners, 
I believe they would come to understand that even the top-down  string
theorists they view as arch-reductionists are, in truth, paragons of 
the emergent impulse. What could be more emergent than the conviction 
that our standard model will arise out of the multiple vacua of M theory and 
the string-theory landscape?

Emergence is, moreover, quite ubiquitous in particle physics, and 
especially in the hadronic physics we are discussing at this meeting. 
For example, as QCD becomes strongly coupled at low energies, new 
phenomena emerge that are not immediately obvious from the 
Lagrangian. Confinement and chiral symmetry breaking, with the implied
appearance of Goldstone bosons, are specific 
illustrations. A graceful description entails new degrees of freedom 
that may be expressed in a model or---in the best of cases---in a new 
effective field theory.

The other hallmark of the kind of science that Pines and Laughlin advocate 
is the synthesis of principles through experiment. This too is central 
to the way hadronic physics is constructed, and runs through the 
agenda of this workshop. So I think that these earnest scholars are 
promoting a combat that need not exist. For 
my part, I can't imagine creating physics without both reduction and 
synthesis, and I think it would be foolish to follow only one 
approach.\footnote{It might be interesting to muse on whether seeing a given 
set of phenomena as emergent represents a transient stage in our evolving 
understanding or a final state. Such considerations call for 
a highly cerebral bottle of wine.}

I hope you will take time, during this meeting and beyond, to reflect 
on  the roles of microscopic parameters and emergent 
phenomena in hadronic physics and in your own work. The topics on the 
agenda lend themselves to building a bridge between the spareness of the 
Lagrangian and the richness of emergent behavior; they can benefit 
from a range of approaches, and they should stimulate rewarding 
conversations among physicists with different sensibilities.

\subsection{Making Connections; Controlling Approximations 
\label{subsec:connect}}
The essence of doing science consists in \textit{making connections} 
that lead us beyond independent explanations for distinct phenoma 
toward a coherent understanding of many phenomena. A network of 
understanding helps us see how different observations fit together 
and---very important---helps us know enough to recognize that 
something \textit{doesn't fit.}

Connections among experiments or observations are not the only
important ones.  Whenever it is possible, we need to make connections
between our models and the QCD Lagrangian---either directly, or through
effective field theories, lattice field theory, or a controlled
approximation to full QCD. I would also stress the potential value of 
reaching toward connections with our knowledge of nuclear forces and 
with the phenomena that occur in nuclear matter under unusal conditions.

My colleague Estia
Eichten~\cite{estiabeijing} likes to emphasize the different
circumstances under which various approximations to QCD can be regarded
as controlled expansions in small parameters.
Nonrelativistic QCD applies to heavy-heavy ($Q_{1}\bar{Q}_{2}$) mesons, for which the
quark masses greatly exceed the QCD scale parameter, $m_{Q_{i}} \gg
\Lambda_{\mathrm{QCD}}$.  Befitting its aptness for the nonrelativistic
limit, NRQCD takes as its expansion parameter $v/c$, the heavy-quark
velocity divided by the speed of light.  Heavy-quark effective theory 
(HQET)
applies usefully to heavy-light ($Q\bar{q}$) systems, for which $m_{Q}
\gg \Lambda_{\mathrm{QCD}}$.  In first approximation, the spin of the
heavy quark is regarded as static, so the ``light-quark spin''
$\vec{\jmath}_{q} = \vec{L}+ \vec{s}_{q}$ is a good quantum number.  The
relevant expansion parameter is {$\Lambda_{\mathrm{QCD}}/m_{Q}$}.
Chiral symmetry  is a valuable starting point for light quark systems
($q_{1}\bar{q}_{2}$) with $m_{q_{i}} \ll \Lambda_{\mathrm{QCD}}$. In 
this case, the expansion parameter compares the current-quark mass to the scale 
of chiral-symmetry breaking, and is generally taken as 
$m_{q}/4\pi f_{\pi}$, where $f_{\pi}$ is the pion 
decay constant. 

\section{Searching for the Relevant Degrees of Freedom 
\label{sec:diquarks}}
Much of our insight into the comportment of hadrons follows from the 
simplifying assumption that mesons are quark--antiquark states, 
baryons are three-quark states, and that the quarks have only 
essential correlations. In the case of baryons, this reasoning leads 
us to the plausible starting point of SU(6)  (flavor-spin) wave functions, 
which indeed offer a useful framework for discussing magnetic moments 
and other static properties. Some observations, however, show us the 
limitations of the zeroth-order guess. If we examine deeply inelastic 
scattering in the limit as $x \to 1$, {spin asymmetries} indicate 
that the SU(6) wave functions are inadequate~\cite{Hughes:1983kf}, and the ratio\footnote{For a clear exposition of issues related to 
the neutron-proton ratio, see Ref.~\cite{Melnitchouk:1995fc}.} 
$F_{2}^{n}/F_{2}^{p}$ is far from the uncorrelated expectation of 
$\cfrac{2}{3}$.

Under what circumstances might it be fruitful---or even essential---to
consider diquarks as physical objects~\cite{Anselmino:1992vg}?  The
algebra of SU(3)$_{c}$ tells us that the $\mathbf{3} \otimes
\mathbf{3}$ quark--quark combination is attractive in the
$\mathbf{3^{*}}$ representation that corresponds to an antisymmetric
diquark structure.  A simple analysis suggests that the attraction of
$[qq]_{\mathbf{3^{*}}}$ is half as strong as that of the
$[q\bar{q}]_{\mathbf{1}}$ ($\mathbf{3} \otimes \mathbf{3^{*}} \to
\mathbf{1}$) channel.  For many years, it has seemed to make sense to
regard members of the scalar nonet \{$f_{0}(600) = \sigma, \kappa(900),
f_{0}(980), a_{0}(980)$\} as $qq\bar{q}\bar{q}$ states organized as
$[[qq]_{\mathbf{3^{*}}}[\bar{q}\bar{q}]_{\mathbf{3}}]_{\mathbf{1}}$~\cite{Jaffe:1976ig}.
Recently, \textit{intrinsic diquarks} ($|uuudc \bar c\rangle$) and
intrinsic double-charm Fock states ($|uud c \bar c c \bar c\rangle$)
have been advanced as an explanation of the production of the SELEX
$\Xi(ccd)$ and $\Xi(ccu)$ states~\cite{Brodsky:2004cv}.  Diquarks as
objects have elicited new attention under the stimulus of experimental
evidence for pentaquark states~\cite{Jaffe:2003sg,shuryak,aneesh}.  (The
attention to pentaquarks should be seen as part of a broader investigation into
the existence of configurations beyond $qqq$ and $q\bar{q}$.) That work,
in turn, has led Wilczek and collaborators to revisit the
Chew--Frautschi systematics of $N,\Delta$
resonances~\cite{Wilczek:2004im}, and to assert that it is useful to
view even low-spin, light baryons as $q[qq]_{\mathbf{3^{*}}}$
configurations.  What can lattice QCD tell us about the shape of $qqq$
baryons---both at the lowest spins and at high angular 
momenta~\cite{Cristoforetti:2004kj}?  Can
the quark--diquark picture be reconciled with intuition from the
$1/N_{c}$ expansion~\cite{'tHooft:1973jz,Dashen:1993jt}?

It is worth testing and extending the $q[qq]_{\mathbf{3^{*}}}$ proposal
by considering its implications for doubly heavy ($QQq$) baryons.  The
comparison with heavy-light ($Q\bar{q}$) mesons offers a chance to
calibrate the attractive forces in the $\mathbf{3^{*}}$ and
color-singlet channels.  Similarly, extending studies of the
systematics of $qq \cdot \bar{q}\bar{q}$ states to $Qq \cdot
\bar{Q}\bar{q}$ states should, over the long term, develop and
challenge the way we think about diquarks.  Finally, in heavy-ion
collisions, we should be alert for tests of the utility of diquarks in
color--flavor locking, color superconductivity, and other novel
phenomena.  Tugging the diquark concept this way and that will help
elucidate the value of colorspin~\cite{Jaffe:1976ih} as an organizing
principle for hadron spectroscopy.

\section{Mesons with Beauty and Charm \label{sec:bsubc}}
There is potentially great value to be gained by stretching our models
and calculations beyond the domains in which we first encountered them.
By leaving the comfort zone, we may happen on effects that were
unimportant---or concealed---in the original setting.  An excellent
example is the prospect of extending our descriptions of the
$\psi\;(c\bar{c})$ and $\Upsilon\;(b\bar{b})$ systems to the spectrum
of $B_{c}\;(b\bar{c})$ mesons~\cite{Eichten:1994gt}.  Establishing the
$B_{c}$ ground state in nonleptonic decays---$\pi\jpsi, a_{1}\jpsi$ are
the most promising final states---to pin down the mass with greater
certainty than is possible in the semileptonic $\jpsi\ell\nu$ channel
and beginning to reconstruct some part of the spectrum in $\gamma$ or
$\pi^{+}\pi^{-}$ cascades to the ground state will be an experimental
\textit{tour-de-force.}


Several factors contribute to the theoretical interest in $B_{c}$.  The
$b\bar{c}$ system interpolates between heavy-heavy ($Q\bar{Q}$) and
heavy-light ($Q\bar{q}$) systems.  The unequal-mass kinematics and the
fact that the charmed quark is more relativistic in a $b\bar{c}$ bound
state than in the corresponding $c\bar{c}$ level imply an enhanced 
sensitivity to effects beyond nonrelativistic quantum mechanics.

The new element in $b\bar{c}$ theory is lattice QCD calculations
that include dynamical quarks.  At this meeting, Andreas Kronfeld reports
new predictions~\cite{kronfeld,Allison:2004hy} for the mass of the $b\bar{c}$ ground
state on behalf of a Glasgow--Fermilab subset of the High-Precision QCD
Collaboration.\footnote{See also the talks by Alan Gray on 
bottomonium~\cite{alang} and by Jim Simone on charmonium~\cite{jims}.}  Using quarkonium as a baseline, they quote a
preliminary value, $M_{B_c} = 6304 \pm 3 \pm 3 \pm 11 ^{+0}_{-12}\mev$;
using a heavy-light baseline, they find $M_{B_c} = 6253 \pm 17 \pm 3
\pm 11 ^{+30\mathrm{-}50}_{-0}\mev$.  They are making final studies of
the sensitivity to lattice spacing and the sea-quark mass, and expect
to present a final result soon.

\section{Chiral Symmetry and the $D_{sJ}$ Levels \label{sec:chisym}}
In early 2003, the Babar~\cite{Aubert:2003fg},
CLEO~\cite{Besson:2003cp}, and Belle~\cite{Abe:2003jk} experiments
presented convincing evidence for two new narrow charmed-strange
mesons, $D_{s}(2317) \to D_{s}\pi^{0}$ and $D_{s}(2459) \to
D_{s}^{*}\pi^{0}$.  These states, which are candidates for the $j_{q} =
\cfrac{1}{2}$ $c\bar{s}$ levels with $J^{PC}=0^{++}$ and $1^{++}$, are
curious on several counts.  Their centroid is well below that 
of the corresponding $j_{q} = \cfrac{3}{2}$ states $D_{s1}(2535)$ and 
$D_{s2}^{*}(2572)$, in disagreement with relativistic quark model 
predictions~\cite{DiPierro:2001uu}.\footnote{See Alexey Drutskoy's 
talk~\cite{drutskoy} for an update on Belle's 
observations.} And although we anticipated 
that the $j_{q} = \cfrac{1}{2}$ $D_{s}$ and $B_{s}$ states might be 
narrower than their nonstrange counterparts, because of the limited 
phase space for kaon emission~\cite{Eichten:1994in},  these states are seen in 
isospin-violating decays; the standard decay channels are 
kinematically forbidden. The unexpected properties of these states have 
provoked much discussion~\cite{godfrey,vanBeveren:2003kd,hwang,estia}, including 
speculations that they might be multiquark or molecular states.

Before looking into interpretations of the narrow charmed-strange 
states, let us take a moment to review some elementary points about 
meson taxonomy. Two useful classification schemes are familiar in 
atomic spectroscopy as the $LS$ and $jj$ coupling schemes. Any state 
can be described in any scheme, through appropriate configuration 
mixing, but it is prudent to keep in mind that a choice of basis can 
guide---or maybe misguide---our thinking.

For equal-mass meson systems ($q\bar{q}$ or $Q\bar{Q}$) it is
traditional to couple the orbital angular momentum, $\vec{L}$, with the
total spin of the quark and antiquark, $\vec{S} =
\vec{s}_{q}+\vec{s}_{\bar{q}}$.  This is the standard practice for
light mesons, and is now familiar for the designation of quarkonium
($c\bar{c}$ and $b\bar{b}$) levels.  The good quantum numbers are
then $S$, $L$, and $J$, with $\vec{J} = \vec{L} + \vec{S}$, and we
denote the spin-singlet and spin-triplet levels as \slj{1}{1}{0} --
\slj{3}{1}{1}; \slj{1}{2}{1} -- \slj{3}{2}{0,1,2}; \slj{1}{3}{2} --
\slj{3}{3}{1,2,3}; and, in general, as $\sLj{1}{L}{L}$ --
$\sLj{3}{L}{L-1,L,L+1}$.

In the case of heavy-light ($Q\bar{q}$) mesons, it is suggestive to
couple the difficult-to-flip heavy-quark spin, $\vec{s}_{Q}$, with 
the ``light spin,'' $\vec{j}_{q} = \vec{L}+\vec{s}_{q}$. The good 
quantum numbers are then $L$, $j_{q}$, and $J$, where $\vec{J} = 
\vec{s}_{Q}+\vec{j}_{q}$, and the low-lying levels are
\begin{eqnarray*}
    L= 0: & j_{q} = \cfrac{1}{2}: & \begin{array}{c} 0^{-}\hbox{ - }1^{-} 
    \end{array} \\[2pt] 
    L = 1: & j_{q} = \left\{\begin{array}{c} \cfrac{1}{2}: \\[2pt] \cfrac{3}{2}: 
    \end{array}\right. & \begin{array}{c}
  0^{+}\hbox{ - }1^{+} \\[2pt]  1^{+}\hbox{ - }2^{+} \end{array}  \\
    \hbox{etc.} &  & 
\end{eqnarray*}
In the absence of configuration mixing, this classification implies 
that the $j_{q} = \cfrac{3}{2}$ states will decay only through the 
$d$-wave, and so will be narrow. The $j_{q} = \cfrac{1}{2}$ states, 
for which $s$-wave decay is allowed, will in general be broad.

It makes sense to seek out intermediate cases wherever we can find 
them. We expect, for example, mixed $1^{+}$ levels in the $B_{c} = 
b\bar{c}$ spectrum, but that information is not likely to be in our hands 
soon. A more accessible case might be that of the strange particles 
($s\bar{q}$), for which the $q\bar{q}$-inspired $LS$  
classification has been the standard. Perhaps some unexpected 
insights might come from considering strange mesons as heavy-light
($Q\bar{q}$) states~\cite{Eichten:1993ub}. In any event, it is worth 
asking how infallible is the intuition we derive from regarding $D_{s}$ 
states as heavy-light.

I think the evidence is persuasive that the $D_{sJ}$ levels are 
ordinary $c\bar{s}$ states at lower masses than anticipated, and I 
find it intriguing that these states might give us a window on chiral 
symmetry in a novel setting~\cite{Bardeen:2003kt}. Let us 
suppose that, contrary to standard intuition in light-quark systems, 
chiral symmetry and confinement might coexist in heavy--light mesons. 
Then we would expect to observe chiral supermultiplets: states with  
orbital angular momenta $L, L+1$, but the same value of $j_{q}$. 
Specifically, we should find the paired doublets
    \begin{displaymath}
	\begin{array}{l}
	  j_{q} = \cfrac{1}{2}:  1\mathrm{S}(0^{-},1^{-})\hbox{ and } 
	1\mathrm{P}(0^{+},1^{+});  \\[6pt]
                      j_{q} = \cfrac{3}{2}:  
                   1\mathrm{P}(1^{+},2^{+})\hbox{ and } 
	1\mathrm{D}(1^{-},2^{-}). 
	\end{array}
    \end{displaymath}
Chiral symmetry predicts equal hyperfine splitting in the paired
doublets, $M_{D_{s}(1^{+})} - M_{D_{s}(0^{+})} = M_{D_{s}(1^{-})} -
M_{D_{s}(0^{-})}$, in agreement with what is observed, and so far, 
the predictions for decay rates match experiment.\footnote{For more 
details, see Estia Eichten's talk~\cite{estia}.} In addition to 
confronting chiral symmetry's predictions for the $D_{s}$ and other 
families, we need to ask to what extent the coexistence of chiral 
symmetry and confinement is realized in QCD.

For any interpretation of the $D_{sJ}$ states, it is imperative to
predict what happens in the $B_{s}$ system. Experimenters need not 
wait for the theorists to place their bets. Tracking down the $B_{sJ}$ 
analogues should be a high priority for CDF and D\O!
\section{The Lost Tribes of Charmonium \label{sec:lost}}
One of the liveliest areas of heavy-flavor spectroscopy has been the
renewed exploration of charmonium and the discovery of several new
states~\cite{Quigg:2004vf}.  Two sets of states have been missing for a
long time from our experimental inventory: $c\bar{c}$ states below
$D\bar{D}$ threshold, and unnatural-parity $c\bar{c}$ states that lie
between $D\bar{D}$ and $D\bar{D^{*}}$ thresholds~\cite{Eichten:2002qv}. 
In the first category, we count the 1\slj{1}{2}{1} $h_{c}$, a $J^{PC} = 
1^{+-}$ expected to lie close to the \slj{3}{2}{J} centroid at 
$3525.3\mev$, and the 2\slj{1}{1}{0} $\eta_{c}^{\prime}$, the $J^{PC} = 
0^{-+}$ hyperfine partner of $\psi^{\prime}(3686)$. The 
$\eta_{c}^{\prime}$ is now firmly established, and good evidence for 
the $h_{c}$ is mounting~\cite{Patrignani,kam,tomaradze}. The second 
category includes the $\eta_{c2}$, the 1\slj{1}{3}{2}  $J^{PC} = 2^{-+}$ 
level, and its hyperfine partner, the $J^{PC} = 2^{--}$ 1\slj{3}{3}{2} 
state, $\psi_{2}$. The discovery of $X(3872)$ is the first evidence 
for a narrow state in the interthreshold region. We do not yet know 
whether it is, in fact, a charmonium state.

The motivation for a continuing examination of the charmonium 
spectrum---even before the additional surprises we have heard from 
Steve Olsen~\cite{steve}---is that there are many states to study, we 
know there is more to QCD than potential models embody, and we hope 
to find experimental evidence for that additional richness. New 
experimental tools offer another stimulus: the opportunity to access 
$J^{PC}$ quantum numbers that were, until recently, difficult to reach. The 
discovery of $X(3872)$ has prompted the realization that there may be 
additional narrow $c\bar{c}$ states with allowed, but inhibited, 
decays to open charm, and that these might constitute the beginning of 
a new facet of charmonium 
spectroscopy~\cite{Barnes:2003vb,Eichten:2004uh,ted}. The most 
promising cases are 1\slj{3}{3}{3}, 2\slj{3}{2}{2}, and 1\slj{3}{4}{4}.

At first sight, it seems natural to think of $X(3872)$ as the missing 
\slj{3}{3}{2} level of charmonium, but some pieces of evidence do not 
fit gracefully with that hypothesis. The mass is higher than the 
$3815\mev$ expected in a single-channel potential 
model, and the radiative decays $X(3872) \to \gamma 
\chi_{c1,2}$---expected to be prominent, or even dominant---have not 
been detected. Other charmonium states have been 
explored~\cite{Pakvasa:2003ea,Barnes:2003vb,Eichten:2004uh}, but the 
\slj{3}{3}{3} level that seems to me most promising still requires 
the discovery of a strong radiative transition to $\chi_{c2}$. It is 
important to remark that our expectations for radiative branching 
fractions depend on knowing the decay rate for $X(3872) \to 
\pi^{+}\pi^{-}\jpsi$. The theory of hadronic cascades is primitive, 
and the normalizing rate, $\Gamma(\psi(3770) \to \pi^{+}\pi^{-}\jpsi)$ 
is poorly known. The absence of signals for $\gamma\gamma \to 
X$~\cite{Dobbs:2004di,petez} argues against some off-beat charmonium 
interpretations. Evidence for a large ($\approx 84\%$) prompt $X$ 
signal at the Tevatron~\cite{Bauer:2004bc,gomez} and the similarity 
of $X(3872)$ production characteristics to those of $\psi^{\prime}$ 
in 2-TeV proton--antiproton collisions~\cite{Abazov:2004kp} are in 
accord with the charmonium hypothesis.

The coincidence of $X(3872)$ with the $D^{0}\bar{D}^{*0}$ threshold at
$3871.5\mev$ also invites alternative interpretations, including an
$s$-wave $D\bar{D}^{*}$ threshold cusp in the $1^{++}$
channel~\cite{Bugg:2004rk,Bugg:2004sh}, and deuteron-like ``molecules''
formed by pion exchange between $D^{0}$ and 
$\bar{D}^{*0}$~\cite{Voloshin:2004mh,Tornqvist:2004qy,Swanson:2003tb,Swanson:2004cq,eric}. 
In the favored $1^{++}$ channel, the hadronic cascade must be 
$X(3872) \to (\pi^{+}\pi^{-})_{I=1}\jpsi$. If the observed dipion 
cascade is the tail of $\rho^{0}$, and if Belle's 
$\pi^{+}\pi^{-}\pi^{0}\jpsi$~\cite{steve} represents the tail of 
$\omega\jpsi$, those features would fit nicely with the molecular 
interpretation. Characterizing the dipion through a study of angular 
distributions~\cite{Rosner:2004ac} is an urgent matter for experiment.

Hybrid ($c\bar{c}g$) states might exist, and these have been put
forward as candidates for
$X(3872)$~\cite{Close:2003mb,Li:2004st,petrov}.  However, lattice
calculations predict that the lightest charmonium hybrids should lie in
the neighborhood of  $4.3\gev (J^{PC} = 1^{-+})$, $4.7\gev (0^{+-})$, 
and $4.9\gev (2^{+-})$~\cite{Liao:2002rj}, and BaBar has seen no evidence 
for an enhanced $X \to \eta\jpsi$ decay rate~\cite{Aubert:2004fc}. 

None of these interpretations implies a charged partner for 
$X(3872)$, and none has been seen~\cite{Aubert:2004kq}.

While we have much work to do to pin down the nature of 
$X(3872)$,\footnote{Long lists of experimental and theoretical tasks 
can be found in my BEACH04 lecture~\cite{Quigg:2004vf}.}  I 
want to emphasize that \textit{there are more states to be found.} 
Belle has given us a hint of a narrow state at $3940\mev$ decaying 
into $D\bar{D}^{*}$~\cite{steve,ziegler}. Perhaps it is the 
2\slj{3}{2}{2} charmonium level, perhaps the coincidence with the 
$D_{s}\bar{D}_{s}$ threshold at $3936.2\mev$ is more than 
coincidence?  And then there is Belle's broader (?) $\omega\jpsi$ 
bump near the same mass. If you stare long enough at the 
$\pi^{+}\pi^{-}\jpsi$ discovery spectra for $X(3872)$, you may see 
another, less prominent, peak about $40\mev$ below. Is it there? What 
limits can we set on other $\pi^{+}\pi^{-}\jpsi$ states?

The discovery of the narrow state $X(3872) \to \pi^+\pi^-\jpsi$ gives
charmonium physics a rich and lively puzzle.  We do not yet know what
this state is.  If the most conventional interpretation as a charmonium
state---most plausibly, the 1\slj{3}{3}{2} or 1\slj{3}{3}{3} level---is
confirmed, we will learn important lessons about the influence of
open-charm states on $c\bar{c}$ levels.  Should the charmonium
interpretation not prevail, perhaps $X(3872)$ will herald an entirely
new spectroscopy.  In either event, several new charmonium states
remain to be discovered through their radiative decays or hadronic
transitions to lower $c\bar{c}$ levels.  Another set of $c\bar{c}$
states promises to be observable as narrow structures that decay into
pairs of charmed mesons.  In time, comparing what we learn from this
new exploration of the charmonium spectrum with analogous states in the
$b\bar{b}$ and $b\bar{c}$ families will be rewarding.  For all three
quarkonium families, we need to improve our understanding of hadronic
cascades.  Beyond spectroscopy, we look forward to new insights about
the production of quarkonium states in $B$ decays and hard scattering.
The rapid back-and-forth between theory and experiment is great fun, 
and I look forward to learning many new lessons!

\section{Concluding Observations \label{sec:conc}}
Hadron spectroscopy is rich in opportunities.  Models---disciplined by
principles---are wonderful exploratory tools that can help us to
uncover regularities and surprises.  It is important that
phenomenological studies make contact at every opportunity with
symmetries and with lattice QCD, especially as the incorporation of
dynamical quarks becomes routine. Our goal---it is the goal of all 
science---must be to build coherent networks of understanding, not 
one-off interpretations of data. In both experiment and theory, in 
both exploration and explanation, we profit by tuning between systems 
with similar but not identical characteristics, and by driving models 
beyond their comfort zones.

I see much to be gained from a comparison of the hadronic body plans 
we know: quark--antiquark mesons and three-quark baryons, with the diversity that springs from 
light and heavy quarks. Light-quark mesons, heavy-light 
mesons, and heavy quarkonia call upon different elements of our 
theoretical armamentarium, as do baryons containing 3, 2, 1, or 0 
light quarks---but all are hadrons, and some of what we learn in one 
setting should serve us in another. Do other body plans occur in 
Nature---two-quark--two-antiquark mesons, four-quark--one-antiquark 
baryons, and more? And what lessons might we draw from the behavior of 
hadronic matter under unusual conditions, including those that prevail 
in heavy-ion collisions?

In addition to the specific measurements I have mentioned and that 
others have highlighted in the course of this meeting, I would like to 
underscore the value of broad searches for new mesons and baryons. 
BaBar's discovery of $D_{sJ}$ and Belle's string of observations remind 
us that you don't have to know precisely what you are looking for to 
find something interesting: combining a convenient trigger particle with an
identifiable hadron or two---$(\jpsi\hbox{ or }\Upsilon) + \pi, \pi\pi,
K, K_{S}, p, \Lambda, \gamma, \eta, \omega, \ldots$---can be very
profitable indeed.

In closing, I'd like to return to Helen Quinn's friendly challenge 
and urge you to take those observations to heart in the way you carry 
out your research and in the way you present it to others. In 
experiment and theory alike, let us use our models and our truncated 
versions of QCD to guide our explorations and organize our 
understanding. Let us keep in mind the limitations of our tools as we 
focus on what we can learn of lasting value. Let us, above all,  try to discern where 
the real secrets are hidden.

\ack
It is a pleasure to thank Estia Eichten and Ken Lane for a delightful
and rewarding collaboration, and to recognize Steve Olsen for ongoing
encouragement and experimental stimulation.  I am grateful to Suh Urk
Chung, Andreas Kronfeld, and Wally Melnitchouk for  helpful remarks.  I congratulate the organizers of this 
first
meeting of the APS Topical Group on Hadronic Physics---Ted Barnes,
Steve Godfrey, Alexei Petrov, and Eric Swanson---for a rich and
provocative program of talks.
Fermilab is operated by Universities Research Association Inc.\ under
Contract No.\ DE-AC02-76CH03000 with the U.S.\ Department of Energy. 
\section*{References}
\bibliography{GHP}
\end{document}